\documentclass[sn-mathphys,Numbered]{sn-jnl}% Math and Physical Sciences Reference Style
\usepackage{graphicx}%
\usepackage{multirow}%
\usepackage{amsmath,amssymb,amsfonts}%
\usepackage{amsthm}%
\usepackage{mathrsfs}%
\usepackage[title]{appendix}%
\usepackage{color}
\usepackage{textcomp}%
\usepackage{manyfoot}%
\usepackage{booktabs}%
\usepackage{algorithm}%
\usepackage{algorithmicx}%
\usepackage{algpseudocode}%
\usepackage{listings}%
\usepackage{lscape,pdflscape}
\usepackage{float}

\theoremstyle{thmstyleone}%
\theoremstyle{thmstyletwo}%

\theoremstyle{thmstylethree}%

\usepackage{cleveref}
\Crefname{figure}{Fig.}{Figs.}
\graphicspath{{Figures/}}

\begin{document}
	
	\title[Joint multifractality between GOIs and uncertainties]{Joint multifractality in the cross-correlations between grains \& oilseeds indices and external uncertainties}
	
	\author[1]{\fnm{Ying-Hui} \sur{Shao}}
	
	\author[2]{\fnm{Xing-Lu} \sur{Gao}}

	\author*[3]{\fnm{Yan-Hong} \sur{Yang}}\email{yangyh@shu.edu.cn}
	
	\author*[2,4,5]{\fnm{Wei-Xing} \sur{Zhou}}\email{wxzhou@ecust.edu.cn}
	
	\affil*[1]{\orgdiv{School of Statistics and Information}, \orgname{Shanghai University of International Business and Economics}, \orgaddress{\city{Shanghai}, \postcode{201620}, \country{China}}}
	
	\affil[2]{\orgdiv{School of Business}, \orgname{East China University of Science and Technology}, \orgaddress{\city{Shanghai}, \postcode{200237}, \country{China}}}
	
	\affil[3]{\orgdiv{SILC Business School}, \orgname{Shanghai University}, \orgaddress{\city{Shanghai}, \postcode{201899}, \country{China}}}
	
	\affil[4]{\orgdiv{Research Center for Econophysics}, \orgname{East China University of Science and Technology}, \orgaddress{\city{Shanghai}, \postcode{200237}, \country{China}}}
	
	\affil[5]{\orgdiv{School of Mathematics}, \orgname{East China University of Science and Technology}, \orgaddress{\city{Shanghai}, \postcode{200237}, \country{China}}}
	

	
	\abstract{This study investigates the relationships between agricultural spot markets and external uncertainties via the multifractal detrending moving-average cross-correlation analysis (MF-X-DMA). The dataset contains the Grains \& Oilseeds Index (GOI) and its five sub-indices of wheat, maize, soyabeans, rice, and barley. Moreover, we use three uncertainty proxies, namely, economic policy uncertainty (EPU), geopolitical risk (GPR), and volatility Index (VIX). We observe the presence of multifractal cross-correlations between agricultural markets and uncertainties.  
		Further,  statistical tests  show that maize has intrinsic joint multifractality with all the uncertainty proxies, exhibiting a high degree of sensitivity. Additionally, intrinsic multifractality among GOI-GPR, wheat-GPR and soyabeans-VIX is illustrated. However, other series have apparent multifractal cross-correlations with high possibilities. Moreover, our analysis suggests that among the three kinds of external uncertainties, geopolitical risk has a relatively stronger association with grain prices.}

	\keywords{Agricultural market, uncertainty, MF-X-DMA, multitfractal cross-correlation, statistical test}

	\maketitle
	
	\section{Introduction}\label{S1:Introduction}	
	
	The impacts of external uncertainties on economic activities and financial markets have attracted wide attention. Extensive research indicates that investors are vulnerable to market sentiment, policy uncertainty, geopolitical risks and other shocks, which may cause a reduction in asset prices, employment rates, and investment rates \cite{Baker-Bloom-Davis-2016-QJEcon}. Additionally, uncertain shocks can have a substantial impact on foreign currency, stock, commodity markets, and intersectoral systemic risk
	\cite{Badshah-Demirer-Suleman-2019-EnergyEcon,Kurov-Stan-2018-JBankFinanc,Shao-Yang-Zhou-2022-PhysicaA,Fazelabdolabadi-2019-FinancInnov,Zhu-Lin-Deng-Chen-Chevallier-2021-EconModel,Le-Pham-Do-2023-EnergyEcon,Huang-Li-Zhang-Chen-2021-IntRevEconFinanc,Fan-Binnewies-DeSilva-2023-JFuturesMark}.	
	
	Particularly in the agricultural market, the impact of external uncertainties is especially prominent and complex. Responsive to climate conditions, macroeconomic fundamentals, energy markets, supply-demand, and policy shifts, price swings occur regularly in agricultural markets \cite{Huang-Yang-2017-GlobFoodSecur}. Moreover, along with increasing participants in agricultural commodity markets, the linkages among commodity markets as well as the co-movement between commodity and equity markets are becoming stronger \cite{Ma-Ji-Wu-Pan-2021-EnergyEcon,Tang-Xiong-2012-FinancAnalJ}. The rise of agricultural financialization might contribute to the price fluctuations in agricultural commodities, causing uncertainties for economy and livelihood.

			Given that the volatility of grain prices is influenced by a variety of factors, particularly uncertainties,  there has been a growing body of studies on the relationship between agricultural markets and uncertainty \cite{Jiang-Ao-Mo-2023-NAmEconFinanc,Akyildirim-Cepni-Pham-Uddin-2022-EnergyEcon}. Sun et al. investigate the impact of trade policy uncertainty (TPU) on agricultural commodity prices by bootstrapping full- and subsample rolling-window Granger causality tests \cite{Sun-Su-Mirza-Umar-2021-Pac-BasinFinancJ}. They find that TPU has a considerable impact on agricultural prices. Moreover, they also report a positive impact of agricultural prices on TPU. Jo{\"e}ts et al. document that agricultural markets are highly sensitive to the level of macroeconomic uncertainty \cite{Joet-Valerie-2017-EnergyEcon}. Yin and Han reveal that higher volatility of commodity markets improves policy uncertainty \cite {Yin-Han-2014-ApplEconLett}. Bakas and Triantafyllou show that uncertainty has a significant effect on commodity price volatility \cite{Bakas-Triantafyllou-2018-JIntMoneyFinan}. Tiwari et al. find that geopolitical risks (GPRs) negatively influence the strong co-movements between commodity and  market energy markets \cite{Tiwari-Boachie-Suleman-Gupta-2021-Energy}. Gozgor et al. investigate the role of Chicago Board of Options Exchange Market Volatility Index (VIX) and equity market uncertainty (EMU) on volatility spillovers from the crude oil to agricultural commodity markets \cite{Gozgor-Lau-Bilgin-2016-JIntFinancMarkInstMoney}.

	Previous research provides evidence of multifractality of agricultural markets and uncertainties. Stosic et al. report multifractal properties of Brazilian agricultural markets 
	\cite {Stosic-Nejad-Stosic-2020-Fractals}. Wang and Feng investigate the multifractality of CBOT agricultural futures and spot markets \cite{Wang-Feng-2020-JStatMech-TheoryExp}. Gao et al. explore the multifractal nature of agricultural spot markets 	\cite{Gao-Shao-Yang-Zhou-2022-ChaosSolitonsFractals}. They find that maize and barley time series exhibit intrinsic multifractal behavior. Moreover, recent studies indicate that economic policy uncertainty index (EPU) has multifractal nature \cite{Yao-Liu-Ju-2020-PhysicaA,Liu-Ye-Ma-Liu-2017-PhysicaA,Gu-Liu-2022-PhysicaA}. To effectively capture the dynamics between uncertainty and grain prices, it is essential to employ a methodology capable of detecting complex patterns. Additionally, multifractal 
			analysis on financial markets offers new insight for market efficiency measurement, forecasting, risk management, and trading strategy \cite{Jiang-Xie-Zhou-Sornette-2019-RepProgPhys,Erer-Erer-Guengoer-2023-FinancInnov,Oral-Unal-2019-FinancInnov}.

			Due to the fractal nature in commodity markets and uncertainties, a bunch of studies have utilized fractal methods to investigate   the relationships among these markets and their connection with uncertainty. Aslam et al. estimate the cross-correlations between economic policy uncertainty and commodity markets, then  confirm the presence of nonlinear dependency \cite{Aslam--Bibi-Ferreira-2022-ResourPolicy}. 
	Wang et al. study the impact of COVID-19 on the cross-correlations between crude oil and agricultural futures \cite{Wang-Shao-Kim-2020-ChaosSolitonsFractals}. They report that multifractal cross-correlations of all the agricultural futures except orange juice future increased after the COVID-19 pandemic. 
	Feng et al. present the asymmetric multifractal cross-correlations between EPU and agricultural futures prices \cite{Feng-Li-Cao-2022-FluctNoiseLett}.

A variety of methods have been developed for joint multifractality analysis \cite{Jiang-Xie-Zhou-Sornette-2019-RepProgPhys}, including the multifractal cross-correlation analysis (MF-CCA) 
			\cite{Oswiecimka-Drozdz-Forczek-Jadach-Kwapien-2014-PhysRevE}, the multifractal cross-correlation analysis based on the partition function approach (MF-X-PF) \cite{Xie-Jiang-Gu-Xiong-Zhou-2015-NewJPhys}, the joint multifractal analysis based on the structure function approach (MF-X-SF) \cite{Antonia-VanAtta-1975-JFM,Schmitt-Schertzer-Lovejoy-Brunet-1996-EPL}, the multifractal detrended cross-correlation analysis (MF-X-DFA)  \cite{Zhou-2008-PhysRevE}, the multifractal detrending moving-average cross-correlation analysis (MF-X-DMA) \cite{Jiang-Zhou-2011-PhysRevE}, and so on. MF-X-DFA is one of the most widely used methods for joint multifractal analysis, while MF-X-DMA performs comparably to this method, or better in some cases \cite{Jiang-Zhou-2011-PhysRevE}. MF-X-DMA method is  based on the multifractal detrending moving average (MF-DMA), 
			which is widely adopted to investigate the multifractal cross-correlations among multiple time series.
			
			However, the multifractal cross-correlations between uncertainty and agricultural markets have not yet been thoroughly examined. Uncertainty arises from various factors, such as political risks, economic fluctuations, and stock market volatility. Current studies might not fully account for the interaction between these factors and  agricultural markets. Besides, the sources of multifractal cross-correlations between the two factors are still not clear, which is crucial for the determination of intrinsic joint multifractality  \cite{Jiang-Xie-Zhou-Sornette-2019-RepProgPhys}. Motivated by these facts, we apply the MF-X-DMA method \cite{Jiang-Zhou-2011-PhysRevE} to better understand the links between agricultural markets and 
	external uncertainties. MF-X-DMA allows a reliable estimation of multifractal cross-correlations \cite{Jiang-Xie-Zhou-Sornette-2019-RepProgPhys,Shao-Liu-Yang-2023-FluctNoiseLett}. Moreover, we investigate components of  joint multifractality between agricultural commodities and uncertainties.

	This study contributes to the literature in several ways. First, we offer new insights into the relationship between agricultural commodity markets and uncertainties through multifractal analysis. The MF-X-DMA method can accurately uncover the nonlinear interaction patterns between agricultural markets and uncertainties. Furthermore, we examine whether the multifractal cross-correlations between the two factors are intrinsic, providing better knowledge of the underlying dynamics in time series. Thus, we can better understand the interplay between agricultural markets and uncertainties. Second, we compare the joint multifractality among different proxies. The dataset contains six global agricultural price indices and three measures of uncertainty, which include economic policy uncertainty (EPU), geopolitical risk (GPR), and market sentiment (VIX). Using a broad range of proxies provides reliable results and enhances the robustness of our findings. This diverse proxy approach is crucial for investigating how different types of external uncertainties interact with agricultural markets. Third, this study can aid policymakers and investors in enhancing their decision-making processes in a world of growing uncertainty.

	This paper is organized as follows. Section~\ref{S2:DataMethodology} describes the datasets and methodology. Section~\ref{S3:Results} provides the results. Section~\ref{S4:Conclusion} summarizes the paper.
	
	\section{Data and methodology}
	\label{S2:DataMethodology}	
	
	\subsection{data}
	
	\begin{figure}[b]
		\centering	
		\includegraphics[width=1\linewidth]{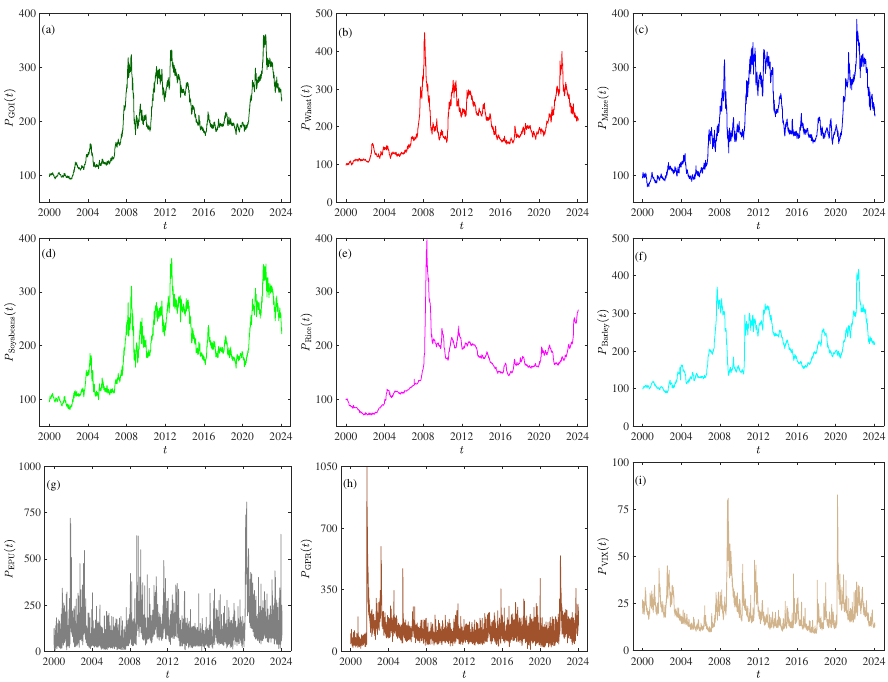}				
		\caption{Historical evolution for daily GOI (a), the wheat price index (b), the maize price index (c), the soyabeans price index (d), the rice price index (e), the barley price index (f), the EPU index (g), the GPR index (h) and the VIX index (i).}
		\label{Fig:Price}
	\end{figure}

	Our data consist of six agricultural price indices from the International Grains Council (IGC), namely, the Grains \& Oilseeds Index (GOI), the wheat price index, the maize price index, the soyabeans price index, the rice price index, and the barley price index. We use the US EPU constructed by Baker et al. \cite{Baker-Bloom-Davis-2016-QJEcon}, GPR proposed by Caldara and Iacoviello \cite{Caldara-Iacoviello-2022-AmEconRev} and VIX as proxies of economic, geopolitical and stock market uncertainty respectively. The dataset covers the period from 4 January 2000 to 29 January 2024 with 6065 daily data points. 
	We choose the period from 2000 to 2024 as our sample time for several reasons. First, this period encompasses several major events, including the 2008 financial crisis, the COVID-19 pandemic, and the Russia-Ukraine war, which had significant impacts on food prices. Second, the choice of a sufficiently long sample period is crucial in ensuring the accuracy of results. If the sample size is too small, it may lead to a narrow scaling range, reflecting inaccurate and misleading analyses of market dynamics 	\cite{Jiang-Xie-Zhou-Sornette-2019-RepProgPhys}. Third, a time span of over 20 years allows researchers to observe long-term trends rather than short-term fluctuations.
	
	Fig.~\ref{Fig:Price} depicts the historical evolution of the GOI index (a), the wheat price index (b), the maize price index (c), the soyabeans price index (d), the rice price index (e), the barley price index (f), EPU (g), GPR (h) and VIX (i). The GOI and the sub-indices exhibited a similar trend, while the three uncertainty measures displayed various behaviors. In 2002, EPU and GPR increased dramatically. The global 2007-08 global financial crisis and food crisis caused a major price surge in the six agricultural price index, EPU and VIX. Between 2010 and 2012, all grains' prices rose, with the exception of rice. During the COVID-19 pandemic, except rice, the grain indices and uncertainties showed a sharp spike.

\begin{figure}[h]
		\centering	
		\includegraphics[width=1\linewidth]{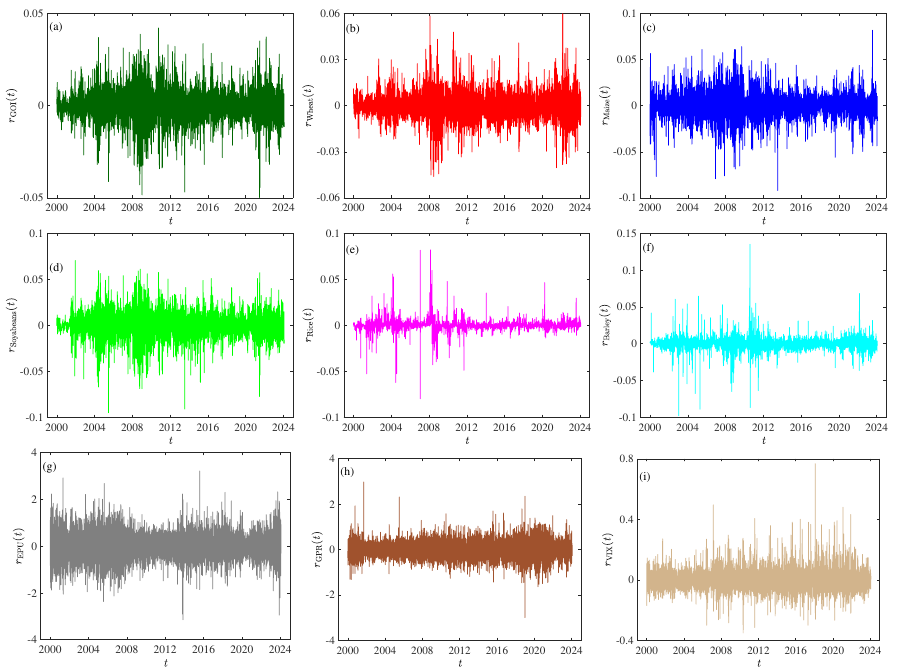}						
		\caption{Daily return series $r(t)$ of agriculture and uncertainty.}
		\label{Fig:Return}
	\end{figure}

	Fig.~\ref{Fig:Return} shows logarithm returns $r(t)$ of the GOI, its five sub-indices and three uncertainty proxies. During the 2007-08 global financial crisis and food crisis, nearly all the grains have seen significant swings.  
	
	\subsection{Cross-correlation test}

	We use a method proposed by Podobnik et al \cite{Podobnik-Grosse-Horvatic-Ilic-Ivanov-Stanley-2009-EurPhysJB} to test whether the cross-correlations between two variables are statistical significant. For time series ${x_{i}}$ and  ${y_{i}}$ with equal length, the statistic is defined as
	\begin{equation}
	Q_{cc}(m)=N^{2} \sum^{m}_{i=1} \frac{X_{i}^2}{N-i},
	\label{Eq:qcc}
	\end{equation}
	where ${X_{i}^2}$ is the cross-correlation function 
	\begin{equation}
	X_{i}= \frac{ \sum^{N}_{k=i+1} x_{k}y_{k-i} }{\sqrt{\sum^{N}_{k=1} x_{k}^{2} \sum^{N}_{k=1} y_{k}^{2}}}.
	\label{Eq:X2}
	\end{equation}

	The statistic $Q_{cc}(m)$ is approximately $\chi^{2}$ distributed with $m$ degrees of freedom. We calculate $Q_{cc}(m)$ statistic to test the null hypothesis that none of the first $m$ cross-correlation coefficient is different from zero. If value of statistic $Q_{cc}(m)$ exceeds the critical values of $\chi^{2}(m)$, we conclude that the null hypothesis is rejected and the cross-correlaitons between the two time series are statistical significant. Otherwise, the null hypothesis is failed to reject.

	\subsection{Multifractal detrending moving-average cross-correlation analysis}
	The process of MF-X-DMA analysis is based on the following steps \cite{Jiang-Zhou-2011-PhysRevE}. For two time series $ X_i$ and $ Y_i$ of equal length $N$, we divide each into $N_{s} =$ int$\left[N/s\right]$ non-overlapping segments of size $s$. We remove the local trends of segments $\{{X}_v(k)\}_{k=1}^s$ and $\{{Y}_v(k)\}_{k=1}^s$ to obtain the cross-correlation function $F_v(s)$ as follows 
	
	\begin{equation}
	F_v(s) = \frac{1}{s}\sum_{k=1}^{s}\left\lvert X_{v}(k)-\widetilde{X}_v(k)\right\rvert \left \lvert Y_{v}(k)-\widetilde{Y}_v(k)\right \rvert,
	\end{equation} 
where $\widetilde{X}_v(k)$ and $\widetilde{Y}_v(k)$ are the local trends functions of $\{{X}_v(k)\}_{k=1}^s$ and $\{{Y}_v(k)\}_{k=1}^s$ respectively. There are a variety of methods to determine the local trends functions.

	In this paper, we calculate the moving average as the local trends functions, in which the algorithm is called MF-X-DMA. The moving average function $\widetilde{Z}(t)$ of $Z\in\{X,Y\}$ in segment with size $s$ is computed as
	\begin{equation}
	\widetilde{Z}(t)= 
	\frac{1}{s}\sum_{k=-\lfloor (s-1)\theta\rfloor }^{\lceil(s-1)(1-\theta)\rceil} Z(t-k),
	\end{equation}
	where $\theta$ is the position parameter varying from 0 to 1. The moving average $\widetilde{Z}(t)$ is calculated in the range of $[t-\lceil(s-1)(1-\theta)\rceil, t+\lfloor (s-1)\theta\rfloor]$. $\theta$=0  corresponds to the backward moving average, in which $\widetilde{Z}(t)$ is computed over all the past data points of the data. $\widetilde{Z}(t)$ with $\theta$=0.5 is called the centered moving average, which covers half past and half future information in each segment. $\theta$=1 refers to the forward moving average, in which $\widetilde{Z}(t)$ contains the  future data points of the signal.

	Then average the $q$th order cross-correlation function to compute the overall fluctuation function 
			$F_{xy}(q,s)$
			
	\begin{equation}
	\left\{
	\begin{array}{ll}
	F_{xy}(q,s)= \left[
	\frac{1}{N_s}\sum_{v=1}^{N_s} { F_v(s) }   ^{q/2}\right]^{1/q},  q\neq0,\\
	F_{xy}(0,s)={\rm exp}\left[\frac{1}{2 N_s}\sum_{v=1}^{N_s}\ln {F_v(s)}  \right ], q=0.
	\end{array}
	\right.
	\end{equation}

	We change values of segment size $s$ and check power-law relationships between $F_{xy}(q,s)$ and $s$ 
	\begin{equation}
	F_{xy}(q,s)\sim s^{H_{xy}(q)}
	\label{Eq:MFDCCA:Fq_hq}.
	\end{equation}

	The scaling exponent $H_{xy}(q)$ serves as an estimator of the generalized bivariate Hurst exponent or the generalized joint
	Hurst exponent versus $q$, which is used as indicator of joint multifractality in time series. If the $H_{xy}(q)$ curves decrease with order $q$, the cross-correlations between the grains \& oilseeds indices and uncertainties are multifractal. Otherwise, the two variable do not exhibit multifractal cross-correlations. The joint mass scaling exponent $\tau_{xy}(q)$ is defined as
	\begin{equation}
	\tau_{xy}(q)=qH_{xy}(q)-1.
	\label{Eq:MFDCCA_tau}
	\end{equation}

	We use $\tau_{xy}(q)$ to characterize and test the joint multifractal nature. And the joint singularity strength function $\alpha_{xy}(q)$ is defined as	
	\begin{equation}
	\alpha_{xy}(q)=\rm{d}\tau_{xy}(q)/\rm{d}q.
	\label{Eq:MFDCCA_a}
	\end{equation}
	
	The joint multifractal spectrum or singularity spectrum $f_{xy}(\alpha)$ is obtained via	
	\begin{equation}
	f_{xy}(\alpha)=q\alpha_{xy}-\tau_{xy}(q).  
	\label{Eq:MFDCCA_fa}
	\end{equation}
	
	Further, we calculate the difference of joint singularity strength function $\alpha_{xy}(q)$ to determine the width of joint singularity,  or the joint singularity width	
	\begin{equation}
	\Delta\alpha_{xy}=\alpha_{xy}(-\infty)-\alpha_{xy}(+\infty) 
	\triangleq \max\alpha_{xy}-\min\alpha_{xy}.
	\label{Eq:MFDCCA_DA}
	\end{equation} 
	The width of joint singularity $\Delta\alpha_{xy}$ is widely used to measure the strength of  cross-correlations in multifractality \cite{Jiang-Xie-Zhou-Sornette-2019-RepProgPhys}. Time series with bigger $\Delta\alpha_{xy}$ exhibit higher  multifractal cross-correlations.

	\subsection{Multifractal cross-correlations test based on the joint mass scaling exponent} 	
	For time series without multifractal cross-correlations, the generalized bivariate Hurst exponent is a constant, which is independent of the order $q$. Thus the joint mass exponent function has a linear relationship with $q$
	\begin{equation}
	\tau_{xy}(q)=qH_{xy}(q)-1=qH_{xy}(2)-1.  
	\label{Eq:MonoDCCA_tau}
	\end{equation}

	And if there exists multifractal cross-correlations between two time series, the generalized bivariate Hurst exponent $H_{xy}(q)$ declines with order $q$. Moreover, the joint mass exponent $\tau_{xy}(q)$ has a nonlinear relationship with order $q$ \cite{Jiang-Xie-Zhou-Sornette-2019-RepProgPhys}. Following Ref.~ \cite{Gao-Shao-Yang-Zhou-2022-ChaosSolitonsFractals}, we test the relationship between $\tau_{xy}(q)$ and $q$ to check the multifractal cross-correlations between two time series
	\begin{equation}
	\tau_{xy}(q)=a_{0}+a_{1}q+a_{2}q^{2}.  
	\label{Eq:tau_test}
	\end{equation}
	If grains \& oilseeds indices and uncertainties exhibit multifractal cross-correlations, $a_{2}\neq0$.

	\subsection{Statistical test for intrinsic multifractality based on the surrogate data}
	
	Extensive research suggests that much of the apparent multifractality in time series stems from fat-tail distribution and/or long-range correlation \cite{Jiang-Xie-Zhou-Sornette-2019-RepProgPhys,Shao-Xu-Liu-Xu-2021-Fractals,Erer-Erer-Guengoer-2023-FinancInnov}. Except these two factors, the nonlinear correlations are necessary for intrinsic multifractality \cite{Jiang-Xie-Zhou-Sornette-2019-RepProgPhys}. Indeed, the main source of the intrinsic multifractality is nonlinear correlation \cite{Zhou-2009-EPL,Drozdz-Kwapien-Oswiecimka-Rak-2009-EPL,Zhou-2012-ChaosSolitonsFractals,Kwapien-Blasiak-Drozdz-Oswiecimka-2023-PhysRevE}, and statistical tests based on proper null models are required to check if the joint multifractality are apparent or intrinsic \citep{Jiang-Xie-Zhou-Sornette-2019-RepProgPhys,Gao-Shao-Yang-Zhou-2022-ChaosSolitonsFractals,Wang-Gao-Zhou-2023-Fractals}. We generate surrogates of each time series using the iterated amplitude-adjusted Fourier transform (IAAFT) \cite{Schreiber-Schmitz-1996-PhysRevLett,Schreiber-Schmitz-2000-PhysicaD}, which have the same distribution and linear long-term correlations except nonlinear correlation with the original time series. It will enable us to eliminate influence of fat-tail distribution and linear correlations.

	For single time series, if the multifractal strength of IAAFT data is higher than original data, the linear correlation, or a fat tail, or the both can be the source of multifractal properties. In this case, nonlinear correlation contributes little to the multifractality in time series, indicating lack of intrinsic multifractal nature. And if the singularity spectrum of IAAFT data shrinks significantly, becomes very weak, or even contracts into small lots, it indicates that multifractality mainly comes from nonlinear correlation. In other conditions, fat-tail distribution, long-range correlation and nonlinear dynamics are the sources of multifractality in time series. For two time series, one should consider carefully that which time series dominates the multifractal cross-correlations. Inspired by Ref.~\cite{Gao-Shao-Yang-Zhou-2022-ChaosSolitonsFractals}, we perform MF-X-DMA on three types of data (1) IAAFT grains \& oilseeds indices and original uncertainties; (2) original grains \& oilseeds indices and IAAFT uncertainties; (3) IAAFT grains \& oilseeds indices and IAAFT uncertainties. The first and third type of time series destroy the nonlinear correlation in GOIs and uncertainties respectively. The second type of surrogate datas eliminate nonlinear correlation both in grains \& oilseeds indices and uncertainty proxies. Surrogates size of 1000 is large enough to provide stable results	\cite{Gao-Shao-Yang-Zhou-2022-ChaosSolitonsFractals}.

	We calculate the joint singularity width $\Delta{\hat{\alpha}_{xy}}$ for each type of surrogate data, then compare $\Delta\alpha_{xy}$ of original data with $ \Delta{\hat{\alpha}_{xy}}$ of surrogate data. For monofractal cross-correlated time series, $\Delta\alpha_{xy}=0$ \cite{Jiang-Xie-Zhou-Sornette-2019-RepProgPhys}. To check whether the time series are monofractal cross-correlated or not, one can test if $\Delta{\hat{\alpha}_{xy}}$ are larger than $\Delta\alpha_{xy}$  \cite{Gao-Shao-Yang-Zhou-2022-ChaosSolitonsFractals}. We determine the $p$-value as the proportion of surrogate data with  $\Delta{\hat{\alpha}_{xy}}>\Delta\alpha_{xy}$. The hypothesis is that 
	the multifractal cross-correlations in surrogate data are stronger than in original data. Given the fact that intrinsic joint multifractality comes from nonlinear process, surrogates with weakened or destroyed nonlinearity should exhibit no stronger joint multifractality than original time series. If we reject the hypothesis for any type of surrogate data, nonlinear correlations in original data are source of joint multifractality. Hence, the original data may have intrinsic multifractal cross-correlations. 
	
	\section{Results}
	\label{S3:Results}

\begin{figure}[!h]
		\centering
		\includegraphics[width=1\linewidth]{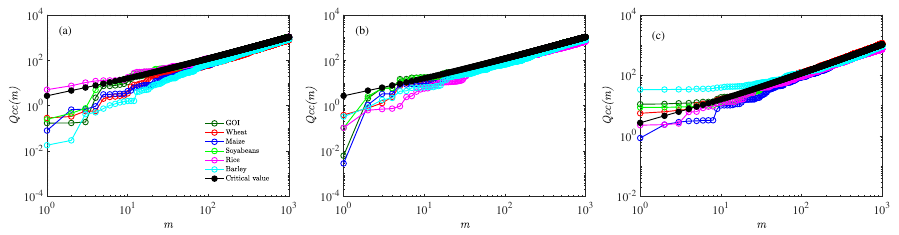}
		\caption{The $Q_{cc}$ statistics between agriculture and EPU (a), GPR (b), and VIX (c).}
		\label{Fig:Qcc}
	\end{figure}

	First we apply the cross-correlation test  \cite{Podobnik-Grosse-Horvatic-Ilic-Ivanov-Stanley-2009-EurPhysJB} to check whether the cross-correlations between the six grains \& oilseeds indices and three uncertainty proxies are statistical significant. We use degrees of freedom $m \in \left[1,1000\right]$ in this paper. Fig.~\ref{Fig:Qcc} displays the log-log plots of  $Q_{cc}(m)$ statistics versus the degree of freedom $m$ between agricultural prices and the three uncertainty measures. The dotted line represents the critical values of $\chi^{2}(m)$ distribution at 5\% significance level. Fig.~\ref{Fig:Qcc} shows that the $Q_{cc}(m)$ statistics of the grains \& oilseeds indices and uncertainties are larger than or close to the critical values of $\chi^{2}(m)$ distribution, which is similar with Ref.~\cite{Jiang-Gu-2016a-PhysicaA}. These results indicate significant cross-correlations between the agricultural price indices and uncertainty measures. 
	
	Then we employ MF-X-DMA to quantify the multifractal cross-correlations between grains \& oilseeds indices and uncertainties. We set $\theta=0$ to eliminate the trend in local segment, which corresponds to the backward moving average.  
	To prevent the finite-size effect and ensure accuracy of measurement, the order $q$ should be properly set \cite{Jiang-Xie-Zhou-Sornette-2019-RepProgPhys}. Too large $q$ might cause fake multifractality into datasets \cite{Jiang-Zhou-2008a-PhysicaA}. One can take $q \in  [-5,5]$ when analyzing time series with thousands of observations. In this paper, each time series consists of 
	6065 data points. Thus we set order $q$ from -5 to 5. Moreover, the range of $s$ is essential for determination of generalized bivariate Hurst exponent, which is estimated by performing a log-log linear regression of fluctuation function $F_{xy}(q,s)$ on segment size $s$. Usually the scaling range spans more than one order of magnitude, which is enough to provide accurate estimation for multifractal analysis. Following Ref.~ \cite{Gao-Shao-Yang-Zhou-2022-ChaosSolitonsFractals}, we set size $s$ from 10 to $10^{2.5}$, which spans 2.5 orders of magnitude.

	Further, we perform statistical test based on the joint mass exponent to check if the multifractal cross-correlations between the six grains \& oilseeds indices and three uncertainty proxies are significant. We regress the the joint mass exponent $\tau_{xy}(q)$ with order $q$ for the GOIs and uncertainties to check if there exists a nonlinear relationship.  
	Then we conduct statistical test based on the surrogate data to investigate if the joint multifractality are intrinsic. We generate 1000 surrogates for each time series using the IAAFT algorithm, compute the joint multifractality of three types of data and compare with original data.

	\subsection{Economic policy uncertainty and IGC price indices}
	
	\begin{figure}[h]
		\centering
		\includegraphics[width=1\linewidth]{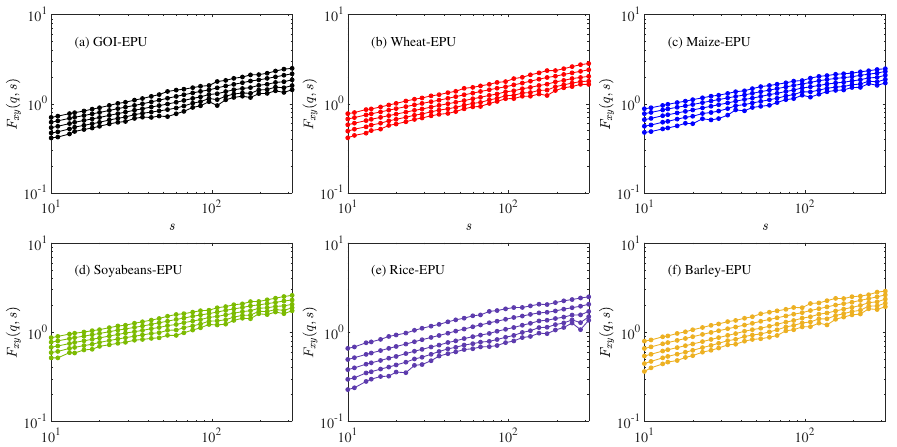}
		\caption{Plots of fluctuation functions $F_{xy}(q,s)$ versus segment size $s$ between the GOI index (a), the wheat price index (b), the maize price index (c), the soyabeans price index (d), the rice price index (e), the barley price index (f) and EPU.}
		\label{Fig:FS:EPU}
	\end{figure}

	\Cref{Fig:FS:EPU} illustrates the log-log plots of MF-X-DMA fluctuation function $F_{xy}(q,s)$ versus segment size $s$ between the grains \& oilseeds indices and EPU for $q=-5, -2.5, 0, 2.5, 5$. As shown in Fig.~\ref{Fig:FS:EPU}, nearly all the lines increase linearly with segment size $s$, implying power-law scaling relationship between fluctuation function $F_{xy}(q,s)$ versus size $s$. However, for larger segment size $s$, the slopes of fluctuation functions for GOI-EPU and wheat-EPU do not decrease with order $q$. This fact puts the existence of multifractal cross-correlations into question. 
	
	\begin{figure}[t]
		\includegraphics[width=1\linewidth]{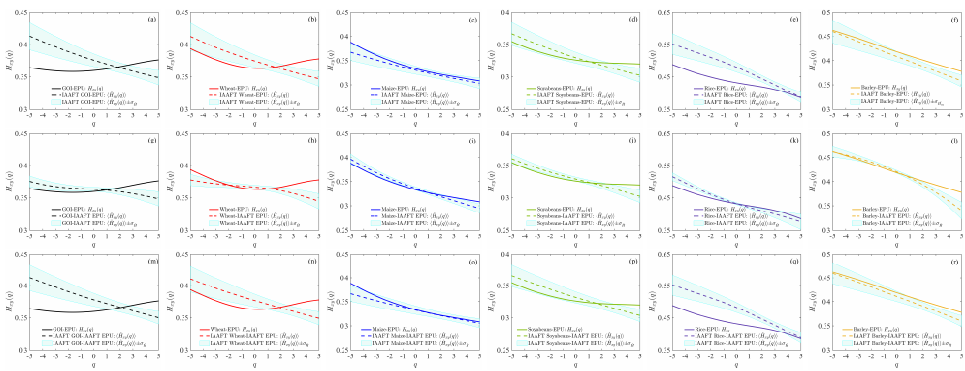}
		\caption{Generalized bivariate Hurst exponent $H_{xy}(q)$ between the IGC indices and EPU. For each series we generate 1000 surrogates with the IAAFT algorithm. We calculate the mean $\langle{\hat{H}_{xy}(q)\rangle}$ and standard deviation $\sigma_{\hat{H}}$ of three types of data, namely IAAFT grains \& oilseeds indices with original EPU, original grains \& oilseeds indices with IAAFT EPU, and IAAFT grains \& oilseeds indices with IAAFT EPU.} 
		\label{Fig:HEPU}
	\end{figure}

	\begin{figure}[b]
		\centering
	\includegraphics[width=1\linewidth]{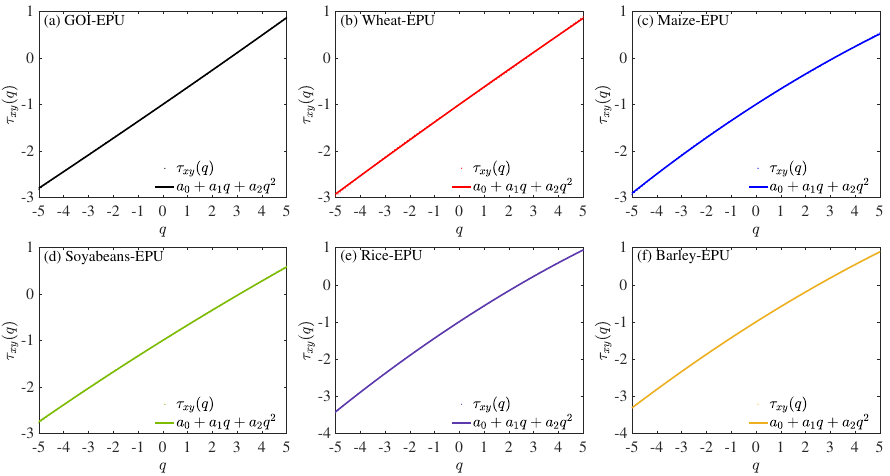}
		\caption{The joint mass exponent $\tau_{xy}(q)$ and polynomial fitting curves for the agricultural price indices and EPU.}
		\label{Fig:FitTauE}
	\end{figure}
	\Cref{Fig:HEPU} shows the generalized bivariate Hurst exponent $H_{xy}(q)$. And in Fig.~\ref{Fig:FitTauE}, the dotted line represents the joint mass exponent $\tau_{xy}(q)$. As seen from \Cref{Fig:HEPU,Fig:FitTauE}, 
	except GOI and wheat, other four grains \& oilseeds indices and EPU have decreasing $H_{xy}(q)$ curves with order $q$ and nonlinear $\tau_{xy}(q)$. However, there is a very small range for the variation of all the $H_{xy}(q)$, which does not exceed 0.1. These facts show that the joint multifractality between GOIs and EPU is very weak.

	\begin{figure}[t]
		\centering
	\includegraphics[width=1\linewidth]{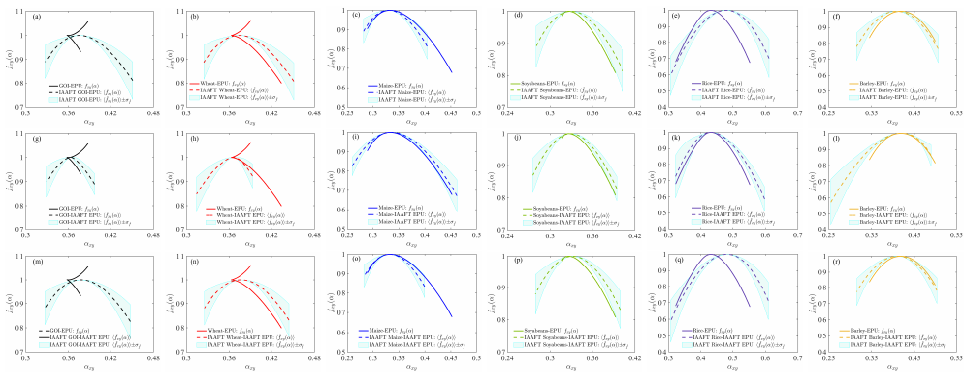}
		\caption{Joint multifractal spectrum $f_{xy}(\alpha)$ versus joint singularity function $\alpha_{xy}$ for the IGC indices and EPU. For each time series, we generate 1000 surrogates using the IAAFT algorithm, then calculate the average $\langle\hat{f}_{xy}(\alpha)\rangle$ and standard variation $\sigma_{\hat{f}}$ of three types of data.}
		\label{Fig:fe}
	\end{figure}

	\begin{table}[b]
		\centering
		\caption{The joint singularity width for the grains \& oilseeds indices and EPU.}
		\label{Tab:MFtest:dalphaE}
		\smallskip
		\setlength\tabcolsep{6pt}
		\begin{tabular}{ccccccccc}
			\toprule
			Pairs  & $\Delta\alpha_{xy}$ & $\langle\Delta\hat{\alpha}_{xy}\rangle$ &  $\sigma_{\Delta\hat{\alpha}}$ &  $p$-value  \\	\midrule

			(IAAFT-GOI, EPU)&0.0300&0.1255&0.0426&0.9880\\
			(GOI, IAAFT-EPU)&           &0.0701&0.0263&0.9440\\
			(IAAFT-GOI, IAAFT-EPU)&&0.1222&0.0399&0.9930\\
						\midrule
			(IAAFT-Wheat, EPU)&0.0706&0.1278&0.0441&0.8910\\
			(Wheat, IAAFT-EPU)&           &0.0795&0.0292&0.6020\\
			(IAAFT-Wheat, IAAFT-EPU)&&0.1198&0.0412&0.8750\\
						\midrule
			(IAAFT-Maize, EPU)&0.1627&0.1237&0.0372&0.1440\\
			(Maize, IAAFT-EPU)&           &0.2023&0.0325&0.8990\\
			(IAAFT-Maize, IAAFT-EPU)&&0.1179&0.0333&0.0870\\
						\midrule
			(IAAFT-Soyabeans, EPU)&0.0763&0.1240&0.0385&0.8860\\
			(Soyabeans, IAAFT-EPU)&          &0.1184&0.0249&0.9560\\
			(IAAFT-Soyabeans, IAAFT-EPU)&&0.1202&0.0372&0.8630\\
						\midrule
			(IAAFT-Rice, EPU)&0.2308&0.3057&0.0655&0.8710\\
			(Rice, IAAFT-EPU)&&0.2770&0.0547&0.8100\\
			(IAAFT-Rice, IAAFT-EPU)&&0.3029&0.0673&0.8550\\
						\midrule
			(IAAFT-Barley, EPU)&0.1542&0.1921&0.0504&0.7800\\
			(Barley, IAAFT-EPU)&           &0.2343&0.0514&0.9750\\
			(IAAFT-Barley, IAAFT-EPU)&&0.1885&0.0490&0.7750\\
			
			\bottomrule
		\end{tabular}
	\end{table}
	
	\Cref{Fig:fe} illustrate the joint singularity spectrum $f_{xy}(\alpha)$. As shown in \Cref{Fig:fe}, apart from GOI and wheat, other grains \& oilseeds indices and EPU have bell-shaped joint singularity spectrum $f_{xy}(\alpha)$. 
	
	To compare strength of multifractal cross-correlations between grains \& oilseeds indices and uncertainties, we calculate joint singularity width $\Delta\alpha_{xy}$ using Eq.~(\ref{Eq:MFDCCA_DA}). \Cref{Tab:MFtest:dalphaE} shows the joint singularity width  $\Delta\alpha_{xy}$ of original data. We find that rice and  EPU have the highest joint multifractality with multifractal width of 0.2308. However, GOI, wheat and soyabeans hardly have multifractal cross-correlations with EPU. Their joint singularity width are 0.0300, 0.0706 and 0.0763, respectively.

	Empirical results suggest that except maize, rice and barley, the grains \& oilseeds indices and EPU exhibit weak multifractal cross-correlations. To check if the grains \& oilseeds indices and EPU possess significant multifractal cross-correlations, we test whether there exists quadratic relationship between the joint scaling exponent function $\tau_{xy}(q)$ and order $q$. We  illustrate the results in Fig.~\ref{Fig:FitTauE} and Table~\ref{Tab:MFtest:tau:EPU}, and observe that the quadratic function fits $\tau_{xy}(q)$ well. The values of coefficient $a_{2}$ for all the IGC indices significantly differ from zero, which provides evidence of multifractal cross-correlations between the grains \& oilseeds indices and EPU. However, coefficient $a_{2}$ for GOI-EPU is positive, which is 0.0013. Moreover, values of coefficient $a_{2}$ for wheat-EPU and soyabeans are close to zero, which are equal to -0.0015 and -0.0034 respectively. The results raise doubts about multifractal cross-correlations in these time series, which accord with our earlier observations.

	\begin{table}[t]	
		\caption{Testing the nonlinearities of the joint mass exponent function $\tau_{xy}(q)$ for the IGC indices and EPU.}
		\centering
		\label{Tab:MFtest:tau:EPU}
		\smallskip
		\setlength\tabcolsep{3pt}
		\begin{tabular}{ccccccccccccccccc}
			\toprule
			& \multicolumn{3}{c}{Full model} &&  \multicolumn{3}{c}{Linear term} &&  \multicolumn{3}{c}{Quadratic term} 				\\			
			\cmidrule{2-4}\cmidrule{6-8}\cmidrule{10-12}
			IGC Indices  & $F$-stat &  $p$-value &  $R^2$ &&  $a_1$  &  $t$-stat  &  $p$-value  &&  $a_2$  &  $t$-stat  &  $p$-value  \\
			
			\midrule
		
			GOI&920934&0.0000&0.9999&&0.3668&1357&0.0000&&0.0013&12&0.0000\\
			Wheat&215921&0.0000&0.9998&&0.3785&657&0.0000&&-0.0015&-6&0.0000\\
			Maize&370208&0.0000&0.9999&&0.3427&858&0.0000&&-0.0079&-51&0.0000\\
			Soyabeans&590706&0.0000&0.9999&&0.3330&1086&0.0000&&-0.0034&-28&0.0000\\
			Rice&1705469&0.0000&1.0000&&0.4358&1843&0.0000&&-0.0100&-109&0.0000\\
			Barley&39163458&0.0000&1.0000&&0.4196&8837&0.0000&&-0.0085&-466&0.0000\\
			
			\bottomrule
		\end{tabular}
	\end{table}

	\begin{figure}[b]
		\centering
		\includegraphics[width=1\linewidth]{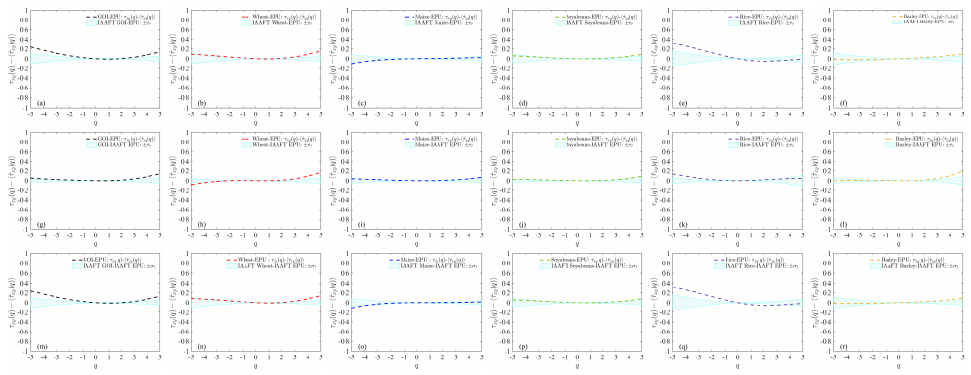}
		\caption{Deviations of the joint mass exponents $\tau_{xy}(q)$ between original IGC indices and EPU from the average joint mass exponents $\langle{\hat{\tau}_{xy}(q)}\rangle$ of the surrogate time series with respect to the order $q$.}
		\label{Fig:DTauEPU}
	\end{figure}
	
	\begin{figure}[t]
		\centering
		\includegraphics[width=1\linewidth]{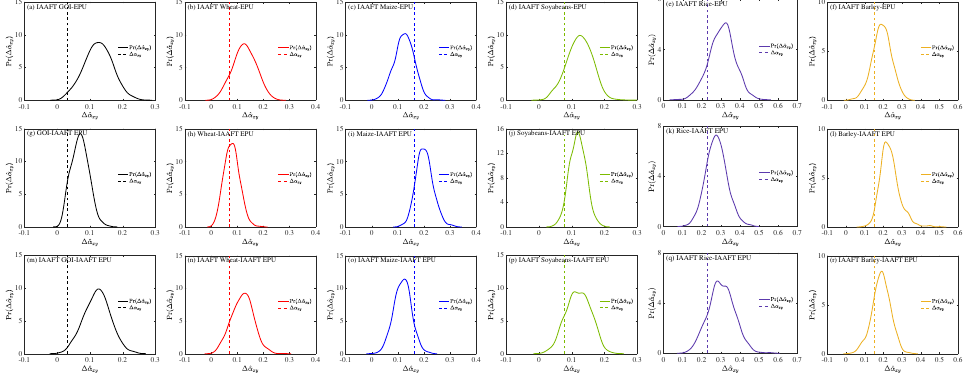}
		\caption{Empirical distribution of the joint singularity widths ${\Delta\hat{\alpha}_{xy}}$ between the IAAFT surrogates of the IGC indices and EPU. The dotted lines are the corresponding singularity widths ${\Delta\alpha}_{xy}$ of the original data sets.}
		\label{Fig:pdfE}
	\end{figure}

	Further, we conduct statistical test based on IAAFT to determine whether the weak multifractal cross-correlations between the grains \& oilseeds indices and EPU are intrinsic. In \Cref{Fig:HEPU}, the solid line shows the $H_{xy}(q)$ of original data, while the dotted line shows the average of generalized bivariate Hurst exponent of surrogate data, which is represented by $\langle{\hat{H}_{xy}(q)\rangle}$. The shadow zone is characterized with $\langle{\hat{H}_{xy}(q)\rangle}\pm\sigma_{\hat{H}}$. \Cref{Fig:fe} shows the average $\langle\hat{f}_{xy}(\alpha)\rangle$ and standard variation $\sigma_{\hat{f}}$. And the shadow zone is defined by  $\langle{\hat{f}_{xy}(q)\rangle}\pm\sigma_{\hat{f}}$. Similarly, Fig.~\ref{Fig:DTauEPU} depicts the deviations between original and surrogated mass exponents with respect to the order $q$. \Cref{Fig:pdfE}  provides the empirical distribution of the joint singularity widths $\Delta {\hat{\alpha}_{xy}}$. And \Cref{Tab:MFtest:dalphaE} represents the average joint singularity width $\langle\Delta\hat{\alpha}_{xy}\rangle $ for three types of surrogates, the singularity widths standard deviation $\sigma_{\Delta\hat{\alpha}}$ of surrogates, and $p$-value. 
	
	We notice a few phenomena. First, as seen from \Cref{Fig:HEPU,Fig:fe,Fig:DTauEPU}, the first type and the third type surrogate data have remarkably similar shadow zones. Besides that, \Cref{Tab:MFtest:dalphaE} shows the first and third kind of surrogates share similar $\langle\Delta\hat{\alpha}_{xy}\rangle$, $\sigma_{\Delta\hat{\alpha}}$ and $p$-value. The result suggests that compared with the IGC indices, the nonlinear correlations in EPU have less impact on the multifractal cross-correlations. 
	Second, \Cref{Fig:HEPU} displays that the generalized bivariate Hurst exponent for almost all the original IGC indices and original EPU are much smoother than all the surrogate data. However, the generalized joint Hurst exponent for maize and EPU is an exception. Additionally, \Cref{Fig:fe} and \Cref{Tab:MFtest:dalphaE} illustrate that except Maize and EPU, almost all the original joint singularity width $\Delta\alpha_{xy}$ is smaller than the average singularity width $\Delta{\hat{\alpha}_{xy}}$ of surrogate data. These findings indicate weaker multifractal cross-correlations in original data for EPU and the other five IGC indices. 
	Moreover, \Cref{Fig:pdfE} and \Cref{Tab:MFtest:dalphaE} show that virtually all surrogate time series have $p$-value bigger than 0.1. However, the third type of surrogates for maize and EPU has a  $p$-value of 0.0870. This finding supports the work of Ref. \cite{Gao-Shao-Yang-Zhou-2022-ChaosSolitonsFractals}. Hence the apparent multifractal cross-correlations is intrinsic at the significance level of 10\% for maize and EPU.

	Given these facts, we conclude that fat-tail distribution and/or long-range linear correlations are contributory factors to the multifractal cross-correlations between EPU and the other five grains \& oilseeds indices. Except EPU and maize, the two variables do not possess intrinsic multifractal cross-correlations with high possibilities. However, the cross-correlations in maize and EPU are likely to be multifractal.

	\subsection{Geopolitical risk and IGC price indices}

	\begin{figure}[!h]
		\centering
		\includegraphics[width=1\linewidth]{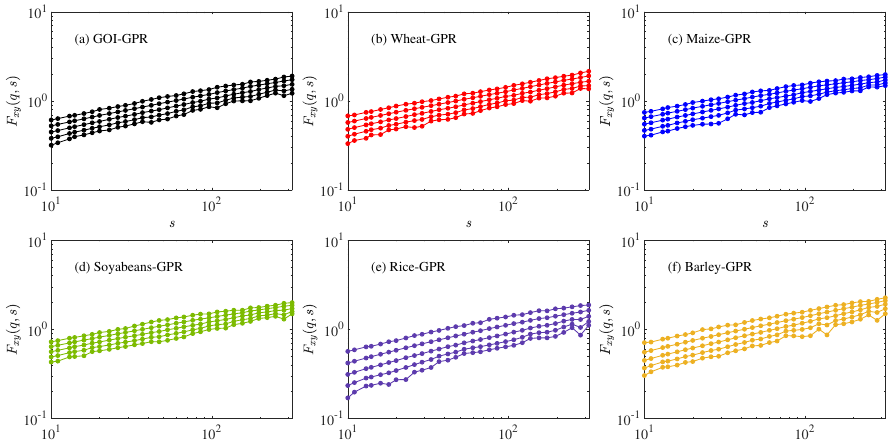}
		\caption{Plots of fluctuation functions $F_{xy}(q,s)$ versus segment size $s$ between the GOI index (a), the wheat price index (b), the maize price index (c), the soyabeans price index (d), the rice price index (e), the barley price index (f) and GPR.}
		\label{Fig:FS:GPR}
	\end{figure}
	
	Fig.~\ref{Fig:FS:GPR} shows the MF-X-DMA fluctuation function $F_{xy}(q,s)$ for the six grains \& oilseeds indices and GPR with $q=-5, -2.5, 0, 2.5, 5$. Nice power-law dependence of the fluctuation functions $F_{xy}(q,s)$ with respect to the scale $s$ are observed. Nearly all the lines rise with the increase of segment size $s$. Moreover, the slopes of fluctuation functions $F_{xy}(q,s)$ increase with order $q$, suggesting the possible existence of multifractal cross-correlations between the grains \& oilseeds indices and GPR.

	\begin{figure}[b]
		\includegraphics[width=1\linewidth]{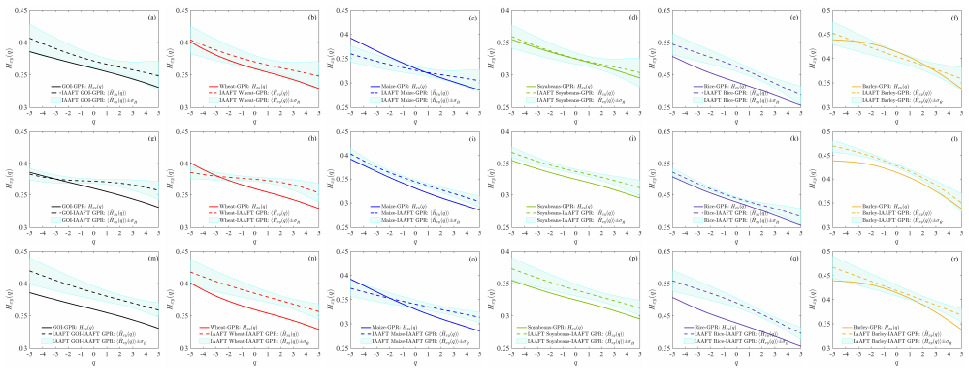}
		\caption{Generalized bivariate Hurst exponent $H_{xy}(q)$ between the IGC indices and GPR. For each series we generate 1000 surrogates with the IAAFT algorithm. We calculate the mean $\langle{\hat{H}_{xy}(q)\rangle}$ and standard deviation $\sigma_{\hat{H}}$ of three types of data, namely IAAFT grains \& oilseeds indices with original GPR, original grains \& oilseeds indices with IAAFT GPR, and IAAFT grains \& oilseeds indices with IAAFT GPR.} 
		\label{Fig:HGPR}
	\end{figure}

	We calculate the value of the generalized bivariate Hurst exponent via Eq.~(\ref{Eq:MFDCCA:Fq_hq}) and illustrate ${H}_{xy}(q)$ in \Cref{Fig:HGPR}. We observe that almost all the $H_{xy}(q)$ decreases with the rise in order $q$, which is consistent with the  fluctuation function $F_{xy}(q,s)$ in Fig.~\ref{Fig:FS:GPR}.
	
\begin{figure}[t]
		\centering	
		\includegraphics[width=1\linewidth]{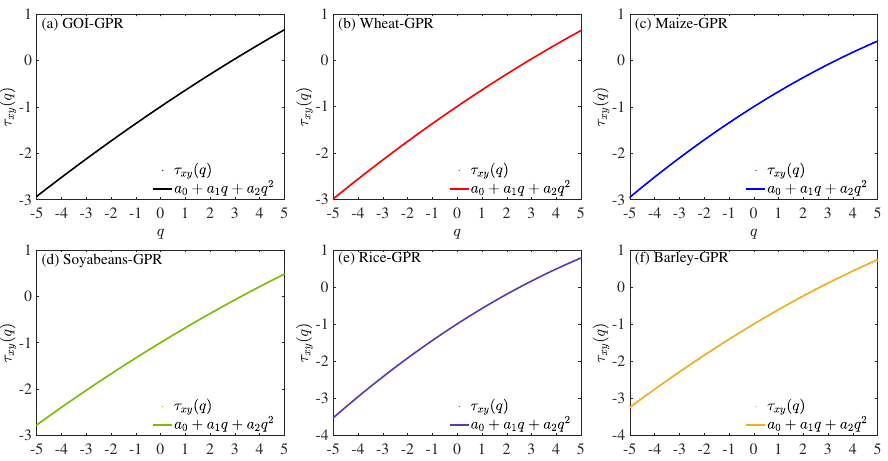}
		\caption{The joint mass exponent $\tau_{xy}(q)$ and polynomial fitting curves for the agricultural price indices and GPR.}
		\label{Fig:FitTauG}
	\end{figure}

	Then we compute the joint mass scaling exponent function $\tau_{xy}(q)$ via the generalized bivariate Hurst exponent  $H_{xy}(q)$. \Cref{Fig:FitTauG} presents the nonlinear relationship between the joint mass scaling exponents $\tau_{xy}(q)$ and order $q$.
	
\begin{figure}[b]
		\centering
		\includegraphics[width=1\linewidth]{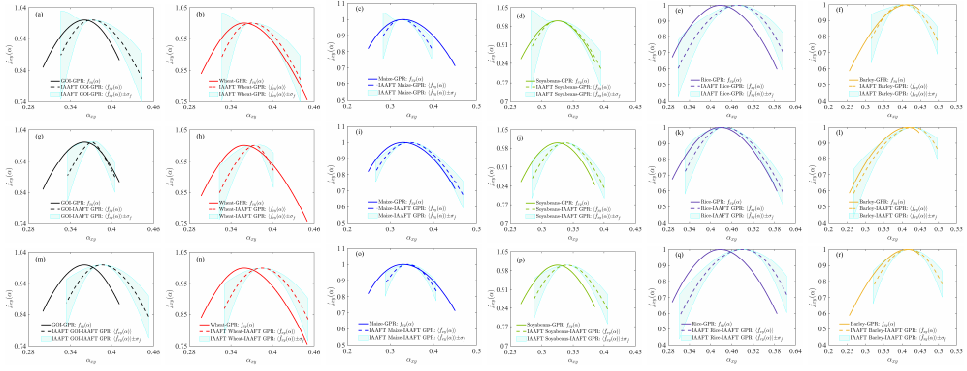}
		\caption{Joint multifractal spectrum $f_{xy}(\alpha)$ versus joint singularity function $\alpha_{xy}$ for the IGC indices and GPR. For each time series, we generate 1000 surrogates using the IAAFT algorithm, then calculate the average $\langle\hat{f}_{xy}(\alpha)\rangle$ and standard variation $\sigma_{\hat{f}}$ of three types of data.}
		\label{Fig:fg}
	\end{figure}

	We calculate the joint singularity strength function and its multifractal spectrum via Eq.~(\ref{Eq:MFDCCA_a}) and Eq.~(\ref{Eq:MFDCCA_fa}). \Cref{Fig:fg} shows that the joint singularity spectrum $f_{xy}(\alpha)$ for all the six grains \& oilseeds indices and GPR have a nice bell shape. These facts suggest presence of joint multifractality in cross-correlations.
	
	\begin{table}[t]
		\centering
		\caption{The joint singularity width for the grains \& oilseeds indices and GPR.}
		\label{Tab:MFtest:dalphaG}
		\smallskip
		\setlength\tabcolsep{6pt}
		\begin{tabular}{ccccccccc}
			\toprule
			Pairs  & $\Delta\alpha_{xy}$ & $\langle\Delta\hat{\alpha}_{xy}\rangle$ &  $\sigma_{\Delta\hat{\alpha}}$ &  $p$-value  \\	\midrule

			(IAAFT-GOI,GPR)&0.1104&0.1259&0.0518&0.5890\\
			(GOI, IAAFT-GPR)&&0.0716&0.0311&0.1070\\
			(IAAFT-GOI, IAAFT-GPR)&&0.1200&0.0409&0.5840\\
			\midrule
			(IAAFT-Wheat, GPR)&0.1531&0.1222&0.0494&0.2590\\
			(Wheat, IAAFT-GPR)&&0.0806&0.0353&0.0260\\
		(IAAFT-Wheat, IAAFT-GPR)&&0.1219&0.0418&0.2300\\
			\midrule
			(IAAFT-Maize, GPR)&0.2002&0.1215&0.0443&0.0390\\
			(Maize, IAAFT-GPR)&&0.2011&0.0364&0.4840\\
		(IAAFT-Maize,IAAFT-GPR)&&0.1187&0.0374&0.0070\\
			\midrule
			(IAAFT-Soyabeans, GPR)&0.1191&0.1216&0.0478&0.4890\\
			(Soyabeans,IAAFT-GPR)&&0.1169&0.0301&0.4590\\	(IAAFT-Soyabeans,IAAFT-GPR)&&0.1188&0.0387&0.5020\\
			\midrule
			(IAAFT-Rice, GPR)&0.2960&0.2973&0.0794&0.5140\\
			(Rice, IAAFT-GPR)&&0.2782&0.0559&0.3860\\
			(IAAFT-Rice,IAAFT-GPR)&&0.3034&0.0703&0.5390\\
			\midrule
			(IAAFT-Barley, GPR)&0.1902&0.1842&0.0600&0.4650\\
			(Barley, IAAFT-GPR)&&0.2297&0.0543&0.7820\\
			(IAAFT-Barley,IAAFT-GPR)&&0.1892&0.0504&0.4840\\
			
			\bottomrule
			
		\end{tabular}
	\end{table}
	
	\Cref{Tab:MFtest:dalphaG} provides the width of singularity spectrum $\Delta\alpha_{xy}(q)$ for the IGC price indices and GPR. GOI-GPR has the smallest singularity spectrum width of 0.1104, followed by soyabeans-GPR, which has singularity spectrum width of 0.1191. Rice-GPR has the highest strength of multifractal cross-correlations, with singularity spectrum width of 0.2960. The results provide evidence that the cross-correlations between the grains \& oilseeds indices and GPR are multifractal. Additionally, compared with EPU, GPR has stronger joint multifractality with all the six the grains \& oilseeds indices.

	\begin{table}[b]
		\centering
		\caption{Testing the nonlinearities of the joint mass exponent function $\tau_{xy}(q)$ for the IGC indices and GPR.}
		\label{Tab:MFtest:tau:GPR}
		\smallskip
		\setlength\tabcolsep{3pt}
		\begin{tabular}{ccccccccccccccccc}
			\toprule
			& \multicolumn{3}{c}{Full model} &&  \multicolumn{3}{c}{Linear term} &&  \multicolumn{3}{c}{Quadratic term} 				\\			
			\cmidrule{2-4}\cmidrule{6-8}\cmidrule{10-12}
			IGC Indices  & $F$-stat &  $p$-value &  $R^2$ &&  $a_1$  &  $t$-stat  &  $p$-value  &&  $a_2$  &  $t$-stat  &  $p$-value  \\
			
			\midrule
			
			GOI&20512012&0.0000&1.0000&&0.3588&6399&0.0000&&-0.0055&-255&0.0000\\
			Wheat&2335866&0.0000&1.0000&&0.3629&2158&0.0000&&-0.0071&-110&0.0000\\
			Maize&1172807&0.0000&1.0000&&0.3359&1526&0.0000&&-0.0107&-127&0.0000\\
			Soyabeans&28417956&0.0000&1.0000&&0.3247&7530&0.0000&&-0.0059&-357&0.0000\\
			Rice&4005693&0.0000&1.0000&&0.4298&2818&0.0000&&-0.0150&-256&0.0000\\
			Barley&164482&0.0000&0.9997&&0.3980&572&0.0000&&-0.0101&-37&0.0000\\		
			\bottomrule
		\end{tabular}
	\end{table}
	
	Next, we test if the joint mass exponents $\tau_{xy}(q)$ has a nonlinear relationship with order $q$ to check the multifractal  cross-correlations between grains \& oilseeds indices and GPR. We regress $\tau_{xy}(q)$ and illustrate the results in Fig.~\ref{Fig:FitTauG} and \Cref{Tab:MFtest:tau:GPR}. We observe that the quadratic polynomial curves fits the joint mass exponents $\tau_{xy}(q)$ well. Table~\ref{Tab:MFtest:tau:GPR} validates the quadratic relationship ($p$-values=0, $R^{2} \approx $ 1, $a_{2}<0$) for all the grains \& oilseeds indices and GPR. These findings accord with the fluctuation function $F_{xy}(q,s)$, generalized bivariate Hurst exponent $H_{xy}(q)$, joint mass exponent function $\tau_{xy}(q)$, and singularity spectrum $f_{xy}(\alpha)$, which report multifractal cross-correlations between the grains \& oilseeds indices and GPR.

	\begin{figure}[t]
		\centering
		\includegraphics[width=1\linewidth]{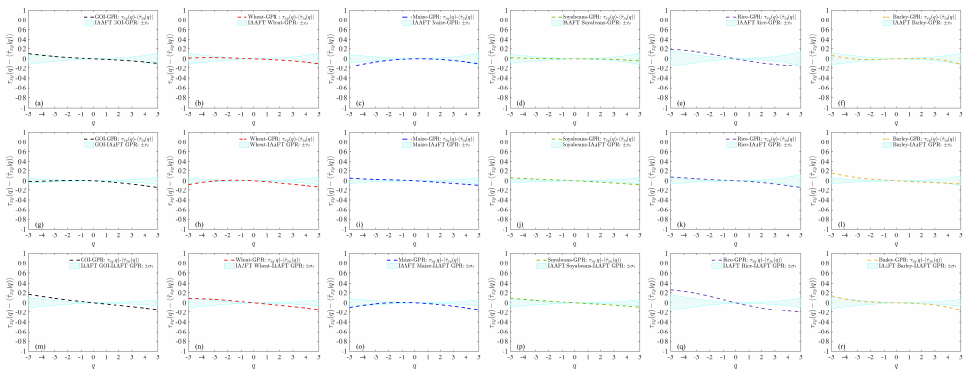}
		\caption{Deviations of the joint mass exponents $\tau_{xy}(q)$ between original IGC indices and GPR from the average joint mass exponents $\langle{\hat{\tau}_{xy}(q)}\rangle$ of the surrogate time series with respect to the order $q$.}
		\label{Fig:DTauGPR}
	\end{figure}
	
\begin{figure}[b]
		\centering
		\includegraphics[width=1\linewidth]{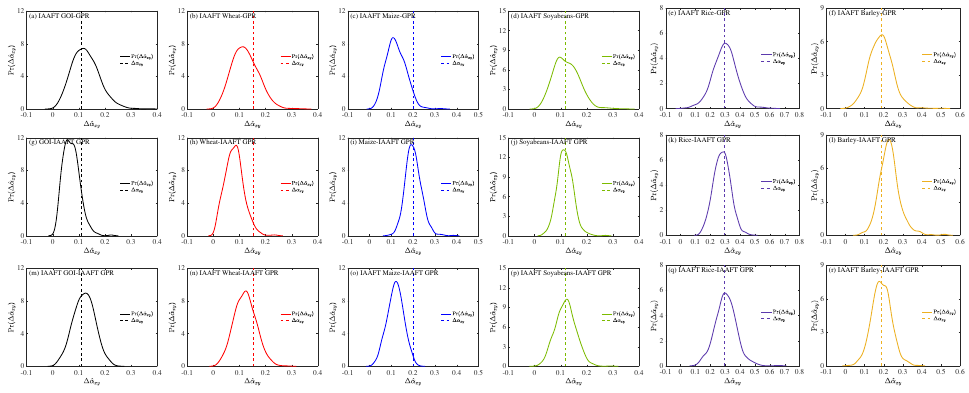}
		\caption{Empirical distribution of the joint singularity widths ${\Delta\hat{\alpha}_{xy}}$ between the IAAFT surrogates of the IGC indices and GPR. The dotted lines are the corresponding singularity widths $\Delta{\alpha}_{xy}$ of the original data sets.}
		\label{Fig:pdfG}
	\end{figure}

	Further, we perform statistical test based on the IAAFT algorithm. We generate three kinds of surrogate data, perform MF-X-DMA analysis and calculate the average. \Cref{Fig:HGPR,Fig:fg,Fig:DTauGPR,Fig:pdfG} show the generalized bivariate Hurst exponent, singularity spectrum, deviations for the joint mass exponent function, and empirical distribution of the joint singularity widths for surrogates respectively. 
	
	As shown in \Cref{Fig:HGPR}, the $\langle{\hat{H}_{xy}(q)\rangle}$ functions are not constant over the range of order $q$, which indicates multifractal cross-correlations in surrogate data. The generalized bivariate Hurst exponent functions of the original IGC indices and original GPR are slightly sharper than those of surrogate data. And we find that the first and third surrogates have similar properties.

	Additionally, as shown in \Cref{Fig:fg,Fig:pdfG} and  \Cref{Tab:MFtest:dalphaG}, second type of surrogate for GOI-GPR with $p$-value of 0.1070 has weaker joint multifractal strength than original data. These variables are possible to have intrinsic joint multifractality in the cross-correlations. 
	And singularity spectrum width of the original wheat-GPR and maize-GPR are broader than that of the surrogate data, or extremely close to it. Both pairs of datasets include surrogates that reject the null hypothesis at significant level of 5\%. These empirical results imply stronger joint multifractality in original data than surrogate time series, which means that besides non Gaussian distribution and/or long-range correlation, embedded nonlinear properties may also contribute to the multifractal cross-correlations between the two grains \& oilseeds indices and GPR.

	However, surrogates of the other three IGC indices fail to reject the null hypothesis of stronger joint multifractal strength, indicating that these variables may not have intrinsic joint multifractality.

	\subsection{VIX and IGC price indices} 
	
\begin{figure}[h]
		\centering
		\includegraphics[width=1\linewidth]{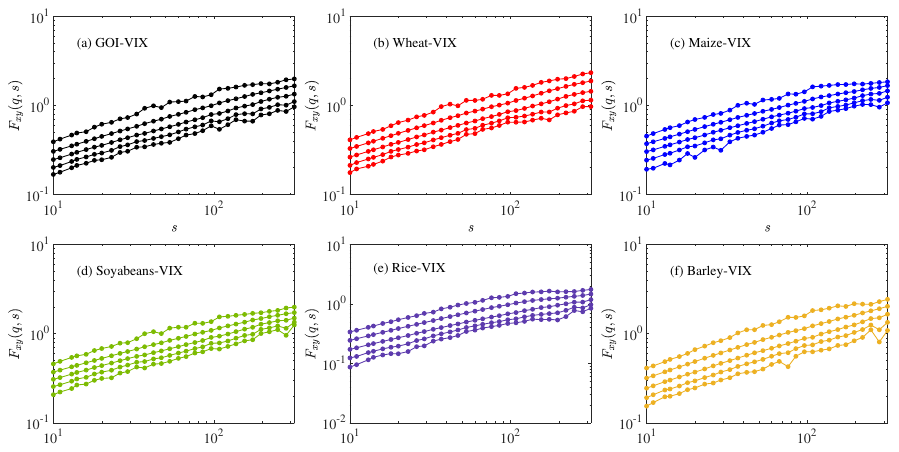}
		\caption{Plots of fluctuation functions $F_{xy}(q,s)$ versus segment size $s$ between the GOI index (a), the wheat price index (b), the maize price index (c), the soyabeans price index (d), the rice price index (e), the barley price index (f) and VIX.}
		\label{Fig:FS:VIX}
	\end{figure}

	Fig.~\ref{Fig:FS:VIX} illustrates the fluctuation functions $F_{xy}(q,s)$ for the grains \& oilseeds indices and VIX with $q=-5, -2.5, 0, 2.5, 5$. Fluctuation functions for maize-VIX, soyabeans-VIX and rice-VIX with $q > 0$ are slightly smoother than those with $q < 0$. 
	
\begin{figure}[t]
		\includegraphics[width=1\linewidth]{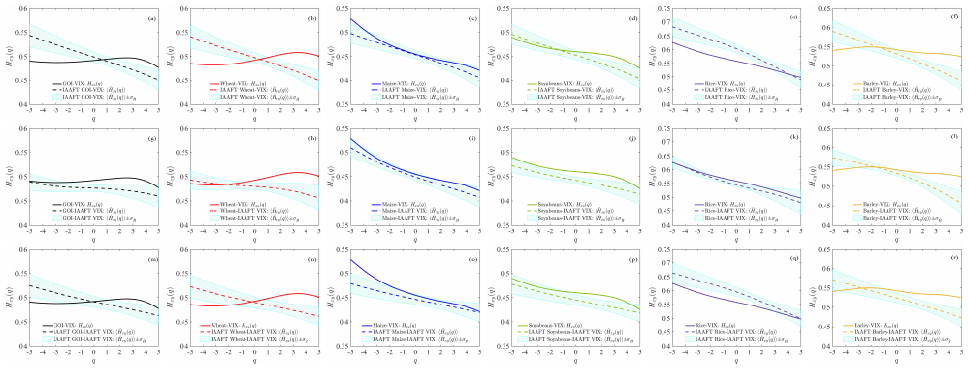}
		\caption{Generalized bivariate Hurst exponent $H_{xy}(q)$ between the IGC indices and VIX. For each series we generate 1000 surrogates with the IAAFT algorithm. We calculate the mean $\langle{\hat{H}_{xy}(q)\rangle}$ and standard deviation $\sigma_{\hat{H}}$ of three types of data, namely IAAFT grains \& oilseeds indices with original VIX, original grains \& oilseeds indices with IAAFT VIX, and IAAFT grains \& oilseeds indices with IAAFT VIX.} 
		\label{Fig:HVIX}
	\end{figure}

\begin{figure}[b]
		\centering
	\includegraphics[width=1\linewidth]{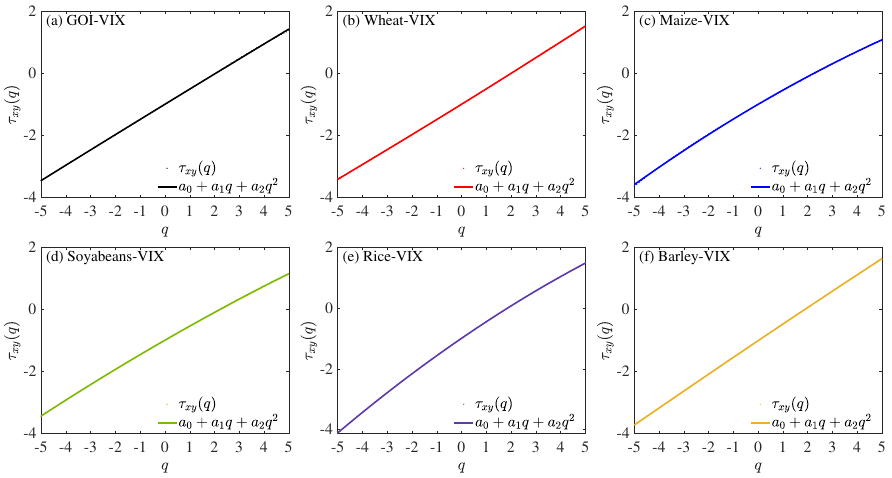}
		\caption{The joint mass exponent $\tau_{xy}(q)$ and polynomial fitting curves for the agricultural price indices and VIX.}
		\label{Fig:FitTauV}
	\end{figure}

\begin{figure}[t]
		\centering
		\includegraphics[width=1\linewidth]{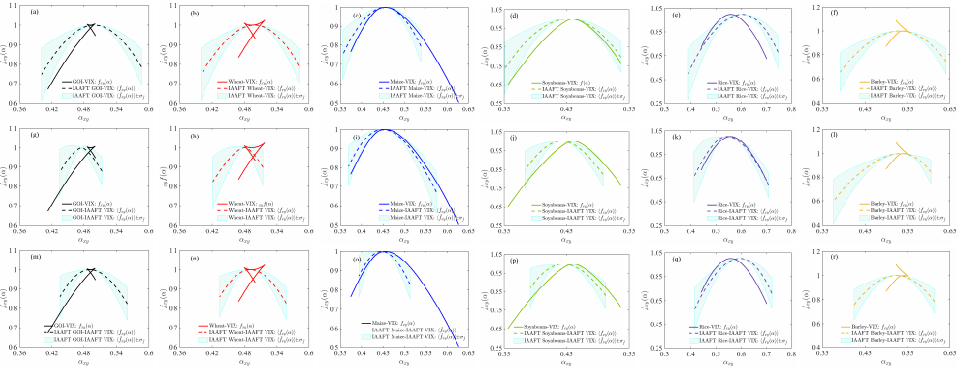}
		\caption{Joint multifractal spectrum $f_{xy}(\alpha)$ versus joint singularity function $\alpha_{xy}$ for the IGC indices and VIX. For each time series, we generate 1000 surrogates using the IAAFT algorithm, then calculate the average $\langle\hat{f}_{xy}(\alpha)\rangle$ and standard variation $\sigma_{\hat{f}}$ for three types of data.}
		\label{Fig:fv}
	\end{figure}

	Moreover, the three time series pairs have a significant decreasing generalized bivariate Hurst index $H_{xy}(q)$ (see in Fig.~\ref{Fig:HVIX}), nonlinear joint mass exponents $\tau_{xy}(q)$ (see in Fig.~\ref{Fig:FitTauV}) and bell-shaped joint singularity spectrum $f_{xy}(\alpha)$ (see in Fig.~\ref{Fig:fv}). These results implies multifractal cross-correlations in maize-VIX, soyabeans-VIX and rice-VIX. However, the other three time series pairs have the monotonic generalized bivariate Hurst exponent function $H_{xy}(q)$, almost linear joint mass exponents $\tau_{xy}(q)$ and knotted joint singularity spectrum $f_{xy}(\alpha)$. These results indicate absence of joint multifractality.

	However, all the $H_{xy}(q)$ varies only in a small range within 0.15. We observe similar results for EPU and GPR. These facts imply that the grains \& oilseeds indices have weak multifractal cross-correlations with uncertainties. \Cref{Tab:MFtest:dalphaV} illustrates that rice and VIX have highest joint singularity width $\Delta\alpha_{xy}$ of 0.2594, while wheat and VIX hardly possess multifractal cross-correlations with joint singularity width of 0.0479. Similarly, GOI-VIX and barley-VIX also hardly exhibit joint multifractality with joint singularity width of 0.0890 and 0.0538, respectively. Moreover, except maize and soyabeans, the other four grains \& oilseeds indices have slightly stronger joint multifractality with GPR than VIX.
	
	\begin{table}[b]
		\centering
		\caption{The joint singularity width for the grains \& oilseeds indices and VIX.}
		\label{Tab:MFtest:dalphaV}
		\smallskip
		\setlength\tabcolsep{6pt}
		\begin{tabular}{ccccccccc}
			\toprule
			Pairs  & $\Delta\alpha_{xy}$ & $\langle\Delta\hat{\alpha}_{xy}\rangle$ &  $\sigma_{\Delta\hat{\alpha}}$ &  $p$-value  \\	\midrule

			(IAAFT-GOI, VIX)&0.0890&0.1874&0.0626&0.9440\\
			(GOI, IAAFT-VIX)&&0.0901&0.0436&0.4700\\
			(IAAFT-GOI, IAAFT-VIX)&&0.1291&0.0515&0.7720\\
			\midrule
			(IAAFT-Wheat,VIX)&0.0479&0.1843&0.0616&0.9920\\
			(Wheat, IAAFT-VIX)&&0.1034&0.0500&0.8700\\
			(IAAFT-Wheat, IAAFT-VIX)&&0.1277&0.0514&0.9430\\
			\midrule
			(IAAFT-Maize, VIX)&0.2523&0.1847&0.0574&0.1080\\
			(Maize, IAAFT-VIX)&&0.2065&0.0552&0.1870\\
			(IAAFT-Maize, IAAFT-VIX)&&0.1235&0.0493&0.0070\\
			\midrule
			(IAAFT-Soyabeans, VIX)&0.1801&0.1861&0.0624&0.5240\\
			(Soyabeans, IAAFT-VIX)&&0.1264&0.0453&0.1320\\
			(IAAFT-Soyabeans, IAAFT-VIX)&&0.1232&0.0478&0.1330\\
			\midrule
			(IAAFT-Rice, VIX)&0.2594&0.3594&0.0938&0.8590\\
			(Rice, IAAFT-VIX)& &0.2980&0.0869&0.6590\\
			(IAAFT-Rice, IAAFT-VIX)&&0.3081&0.0855&0.7250\\
			\midrule
			(IAAFT-Barley, VIX)&0.0538&0.2499&0.0730&0.9970\\
			(Barley, IAAFT-VIX)& &0.2317&0.0751&0.9940\\
			(IAAFT-Barley, IAAFT-VIX)&&0.1910&0.0625&0.9880\\		
			\bottomrule
		\end{tabular}
	\end{table}

	\begin{table}[t]
		\centering
		\caption{Testing the nonlinearities of the joint mass exponent function $\tau_{xy}(q)$ for the IGC indices and VIX.}
		\label{Tab:MFtest:tau:VIX}
		\smallskip
		\setlength\tabcolsep{3pt}
		\begin{tabular}{ccccccccccccccccc}
			\toprule
			& \multicolumn{3}{c}{Full model} &&  \multicolumn{3}{c}{Linear term} &&  \multicolumn{3}{c}{Quadratic term} 			\\			
			\cmidrule{2-4}\cmidrule{6-8}\cmidrule{10-12}
			IGC Indices  & $F$-stat &  $p$-value &  $R^2$ &&  $a_1$  &  $t$-stat  &  $p$-value  &&  $a_2$  &  $t$-stat  &  $p$-value  \\
			\midrule					
				
			GOI&662875.0064&0.0000&0.9999&&0.4893&1151&0.0000&&-0.0008&-4&0.0000\\
			Wheat&1342585.1051&0.0000&1.0000&&0.4947&1638&0.0000&&0.0020&17&0.0000\\
			Maize&320505.6429&0.0000&0.9998&&0.4676&799&0.0000&&-0.0102&-45&0.0000\\
			Soyabeans&1065116.0845&0.0000&1.0000&&0.4599&1458&0.0000&&-0.0060&-49&0.0000\\
			Rice&4640183.6284&0.0000&1.0000&&0.5603&3040&0.0000&&-0.0128&-181&0.0000\\
			Barley&1285246.7485&0.0000&1.0000&&0.5377&1603&0.0000&&-0.0015&-11&0.0000\\
			
			\bottomrule
		\end{tabular}
	\end{table}
	
	\begin{figure}[b]
		\centering
		\includegraphics[width=1\linewidth]{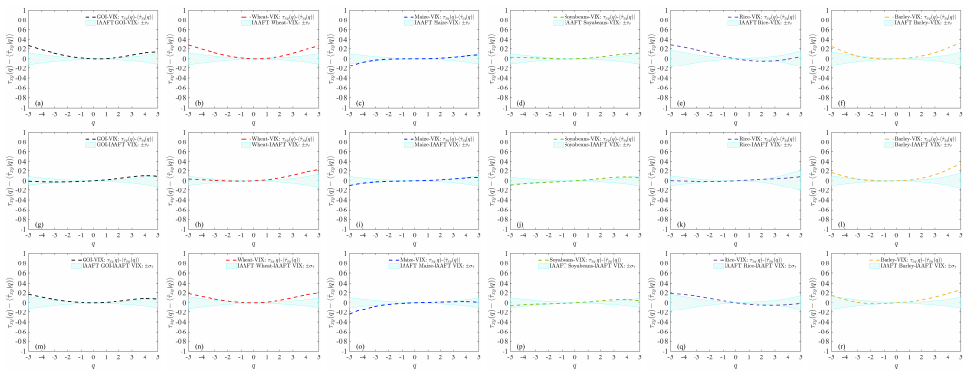}
		\caption{Deviations of the joint mass exponents $\tau_{xy}(q)$ between original IGC indices and VIX from the average joint mass exponents $\langle{\hat{\tau}_{xy}(q)}\rangle$ of the surrogate time series with respect to the order $q$.}
		\label{Fig:DTauVIX}
	\end{figure}
	
	To test if the grains \& oilseeds indices and VIX possess significant multifractal cross-correlations, we regress $\tau_{xy}(q)$ with order $q$. Fig.~\ref{Fig:FitTauV} and Table~\ref{Tab:MFtest:tau:VIX} suggest a quadratic relationship between $\tau_{xy}(q)$ and $q$. 
	Apart from that, the coefficient $a_{2}$ for wheat-VIX is positive, which is equal to 0.0020. Moreover, $a_{2}$ for GOI-VIX and barley-VIX are very small, being -0.0008 and -0.0015, respectively.
	 These facts imply lack of multifractal cross-correlations in these variables. The coefficients $a_{2}$ for other time series are significantly different from zero, which indicates possible existence of multifractal cross-correlations between the other three grains \& oilseeds indices and VIX.
	
\begin{figure}[h]
		\centering
		\includegraphics[width=1\linewidth]{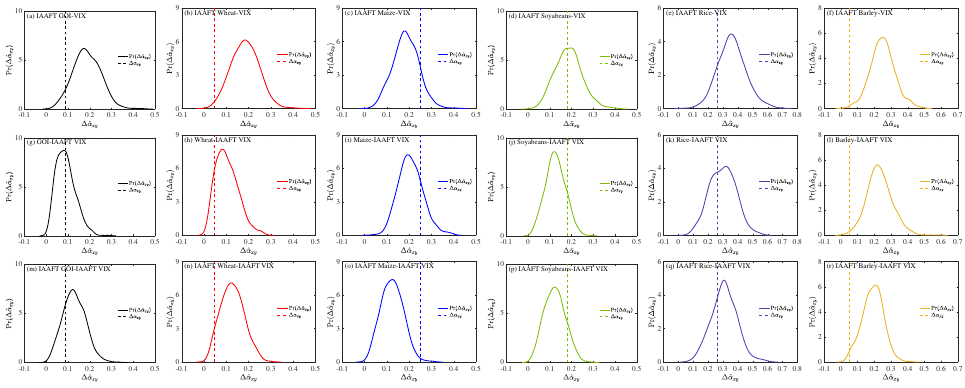}
		\caption{Empirical distribution of the joint singularity widths ${\Delta\hat{\alpha}_{xy}}$ between the IAAFT surrogates of the IGC indices and VIX. The dotted lines are the corresponding singularity widths ${\Delta\alpha}_{xy}$ of the original data sets.}
		\label{Fig:pdfV}
	\end{figure}

	We then perform statistical test based on the IAAFT surrogates and reports results in \Cref{Fig:HVIX,Fig:fv,Fig:DTauVIX,Fig:pdfV} and \Cref{Tab:MFtest:dalphaV}. We observe that the first and third surrogates share similar multifractal features and statistical properties. Moreover, as is shown in \Cref{Fig:HVIX}, except maize, the other original GOIs and VIX have smoother generalized bivariate Hurst exponents than all types of surrogate data. Besides that, \Cref{Fig:fv} and \Cref{Tab:MFtest:dalphaV} suggest that except maize-VIX and soyabeans-VIX, surrogate data have broader singularity width than original data. These facts imply that non Gaussian distributions or/and linear correlations in time series may cause the multifractal cross-correlations in grains \& oilseeds indices and VIX. 
	
	Additionally, \Cref{Fig:pdfV} and \Cref{Tab:MFtest:dalphaV} provide the empirical distribution. Almost all series have $p$-value bigger than 0.1, while third type of surrogate data for maize and VIX rejects the null hypothesis with a $p$-value of 0.0070. And $p$-value of second and third type of surrogate data for soyabeans and VIX is 0.1320 and 0.1330 respectively. Nonlinearity in time series contributes to the joint multifractality in maize-VIX and soyabeans-VIX. We conclude that fat-tail distribution and/or long-range linear correlations are contributory factors to the apparent multifractal cross-correlations between grains \& oilseeds indices and VIX.

	\section{Conclusion}
	\label{S4:Conclusion}

	In this paper, we investigated the multifractal cross-correlations between grain market and uncertainties via MF-X-DMA. We studied daily data of the GOI index and its five sub-indices for wheat, maize, soyabeans, rice and barley as proxies for grain market. The sample spanned from 4 January 2000 to 29 January 2024 with 6065 data points. We considered three uncertainty measurements, namely, EPU, GPR and VIX. We calculated the $Q_{cc}(m)$ statistic to check whether the cross-correlations are statistical significant. Moreover, we performed two statistical tests to explore whether the multifractal cross-correlations are intrinsic or not.

	The $Q_{cc}(m)$ test shows that the cross-correlations between the IGC indices and uncertainties are statistical significant. Multifractal analysis results show that all the $H_{xy}(q)$ varies only in a small range within 0.15, which provides evidence of weak multifractal cross-correlations between most time series. The results corroborate the findings of previous work in the association between agricultural markets and uncertainties \cite{Joet-Valerie-2017-EnergyEcon,Yin-Han-2014-ApplEconLett,Bakas-Triantafyllou-2018-JIntMoneyFinan,Tiwari-Boachie-Suleman-Gupta-2021-Energy,Gozgor-Lau-Bilgin-2016-JIntFinancMarkInstMoney,Huang-Li-Zhang-Chen-2021-IntRevEconFinanc,Fan-Binnewies-DeSilva-2023-JFuturesMark}.  
	However, soyabeans spot might hardly exhibit joint multifractal behavior with EPU, while Feng et. al \cite{Feng-Li-Cao-2022-FluctNoiseLett} report multifractal cross-correlations between soyabean futures and EPU. This inconsistency may be due to the diverse datasets. Futures prices reflect traders' expectations of future market conditions. Traders may anticipate that changes in economic policy uncertainty will impact the prices of soybeans, leading to interactions between the two factors. However, the spot market more closely reflects current conditions, taking longer to absorb and react to policy changes. In addition, rice and barley hardly have intrinsic multifractal cross-correlations with uncertainties.
	
	Moreover, there are hardly any multifractal cross-correlations in GOI-EPU, wheat-EPU, GOI-VIX, wheat VIX and barley-VIX. Additionally, GPR has stronger joint multifractality with all the grains \& oilseeds indices than EPU does. And compared with VIX, GPR also exhibits a stronger correlation with four IGC indices except maize and soyabeans. 
	These results seem to be consistent with other research which found that spillover effect between EPU and agricultural commodity markets are weaker in the long term than short term  \cite{Jiang-Ao-Mo-2023-NAmEconFinanc}, 
	while geopolitical risks significantly affect food prices \cite{Saadaoui-BenJabeur-Goodell-2022-FinancResLett}. Another possible explanation is that the food market is highly globalized with extensive trade. Most countries need to import food and grains to meet needs. Thereby, geopolitical changes in major exporting countries can significantly impact global food prices. However, EPU and VIX measure policy uncertainty and investor sentiment respectively. These factors might be less directly connected to agricultural factors like weather conditions, yields, supply and demand. This may cause weaker joint multifractality in cross-correlations.
	
	Then we performed statistical test based on the joint mass scaling exponent to check if the joint multifractality is statistical significant. We regressed the joint scaling exponent function with order $q$. The coefficient of most variables significantly differs from zero, which provides evidence of multifractal cross-correlations between time series. Howevere, coefficients for GOI-EPU, wheat-EPU, soyabeans-EPU, GOI-VIX, wheat-VIX and barley-VIX are close-to-zero or positive, which suggest lack of joint multifractality. 	
	
	Further, we used statistical tests based on IAAFT to determine the intrinsic multifractal cross-correlations. Statistical tests show that grains \& oilseeds indices have greater impact on the cross-correlations than uncertainty proxies. Among the six agricultural indices, only maize has intrinsic joint multifractality with EPU at the significance level of 10\%. And our work suggests the possible existence of intrinsic joint multifractality in GOI-GPR, maize-GPR and wheat-GPR. Additionally, we observed intrinsic multifractality in maize-VIX and soyabeans-VIX. However, other series exhibit apparent multifractal cross-correlations with high possibilities. In conclusion, we report heterogeneous links between agricultural markets and uncertainties. 
	
	We then offer a more detailed look at the intrinsic joint multifractality among the agricultural indices and external uncertainties. Both GOI, maize and wheat exhibit intrinsic joint multifractality with GPR. It may be that geopolitical risks directly affect the production and distribution of grains. Besides that, maize exhibits intrinsic multifractal cross-correlations with all three uncertainties. The possible reasons are as follows. The demand for maize is very high, making it a major food source globally. Many countries need to import maize from abroad. Any factors affecting the supply chain, such as political conflicts and economic fluctuations, can quickly impact maize prices, leading to a strong correlation with external uncertainties. Moreover, the financialization of maize and soyabeans \cite{Ma-Ji-Wu-Pan-2021-EnergyEcon,Tang-Xiong-2012-FinancAnalJ}, might strengthen the co-movement between the two agricultural indices and stock markets.

Our work has the following implications. First, strengthening the resilience of the agricultural sector against geopolitical shocks is essential for policymakers and investors. Establishing early-warning systems to supervise global events, economic indicators, market trends, and their impacts on food markets is important. Both policymakers and investors should monitor political developments for grains sensitive to geopolitical risks. And maize is highly sensitive to external fluctuations. They should also pay close attention to policy shifts and stock market indicators to address these external fluctuations. Second, policymakers and market participants should be prepared for sudden price volatility in certain commodities and develop strategies to mitigate the impact of uncertainties. Given the high sensitivity of food markets, decision-makers should diversify import sources and establish strategic food reserves to ensure a stable grain supply. Investors should diversify their portfolios to mitigate risks. They could invest in a mix of grains with different risk profiles or explore other asset classes. It's also crucial to seize opportunities in sectors that could benefit from external uncertainties.

	\backmatter
	
	%\bmhead{Supplementary information}
	
	%If your article has accompanying supplementary file/s please state so here. 

	%\section*{Ethics approval and consent to participate}%% if any
	%Text for this section\ldots
	
	%\section*{Consent for publication}%% if any
	%Text for this section\ldots

	\section*{Declarations}
	
	%Some journals require declarations to be submitted in a standardised format. Please check the Instructions for Authors of the journal to which you are submitting to see if you need to complete this section. If yes, your manuscript must contain the following sections under the heading `Declarations':
	
	%\begin{itemize}
	%\item Conflict of interest/Competing interests (check journal-specific guidelines for which heading to use)
	%\item Ethics approval 
	%\item Consent for publication
	%\item Availability of data and materials
	%\item Authors' contributions
	%\end{itemize}
	
	%\noindent
	%If any of the sections are not relevant to your manuscript, please include the heading and write `Not applicable' for that section. 
	
	%\bigskip
	%\bigskip\noindent
	%Springer journals and proceedings: \url{https://www.springer.com/gp/editorial-policies}
	%
	%\bigskip\noindent
	%Nature Portfolio journals: \url{https://www.nature.com/nature-research/editorial-policies}
	%
	%\bigskip\noindent
	%\textit{Scientific Reports}: \url{https://www.nature.com/srep/journal-policies/editorial-policies}
	%
	%\bigskip\noindent
	%BMC journals: \url{https://www.biomedcentral.com/getpublished/editorial-policies}
	%\end{flushleft}
	
	\bmhead{Availability of data and materials}%% if any
	The datasets are available from the following sources. Grains \& oilseeds indices: the International Grains Council; EPU: http://www.policyuncertainty.com; GPR: https://www.matteoiacoviello.com/gpr.htm; VIX: Chicago Board Options Exchange. 
	
	\bmhead{Competing interests}
	The authors declare that they have no competing interests.

	%	\bmhead{Authors' contributions}
	%	% 	This work was conducted in collaboration with all authors. All authors read and approved the final manuscript.
	%	...
	%	
	%	
	%	\bmhead{Acknowledgments}
	%	% We sincerely thank all the editors and reviewers for their insightful comments.
	%	...
	%	
	\bmhead{Abbreviations}
	DMA: Detrending moving average; EMU: Equity market uncertainty; EPU: Economic policy uncertainty; GOI: Grains \& Oilseeds Index; GPR: Geopolitical risk; IAAFT: Iterated amplitude-adjusted Fourier transform; IGC: International Grains Council; MF- X-DMA: multifractal detrending moving-average cross-correlation analysis; VIX: Volatility Index.

\end{document}